\documentclass[12pt]{iopart}
\usepackage{graphicx}
\usepackage{epsfig}
\usepackage{array}
\begin{document}

\title{Spin dynamics in the optical cycle of single nitrogen-vacancy centres in diamond}

\author{Lucio Robledo, Hannes Bernien, Toeno van der Sar and Ronald Hanson}

\address{Kavli Institute of Nanoscience Delft, Delft University of Technology, P.O. Box 5046, 2600 GA Delft, The Netherlands}
\ead{l.m.robledoesparza@tudelft.nl}
\begin{abstract}
We investigate spin-dependent decay and intersystem crossing in the optical cycle of single negatively-charged nitrogen-vacancy (NV) centres in diamond. We use spin control and pulsed optical excitation to extract both the spin-resolved lifetimes of the excited states and the degree of optically-induced spin polarization. By optically exciting the centre with a series of picosecond pulses, we determine the spin-flip probabilities per optical cycle, as well as the spin-dependent probability for intersystem crossing. This information, together with the indepedently measured decay rate of singlet population provides a full description of spin dynamics in the optical cycle of NV centres. The temperature dependence of the singlet population decay rate provides information on the number of singlet states involved in the optical cycle.
\end{abstract}

\maketitle

\section{Introduction}
Nitrogen-vacancy (NV) centres in diamond are well-defined quantum systems in the solid state, with excellent spin coherence properties \cite{Balasubramanian09}. Even at ambient conditions, NV centres have successfully been used in the field of quantum information processing \cite{Jelezko04b, Childress06, Dutt07, Neumann08, Fuchs09, Neumann10, Gijs}, magnetic sensing \cite{Degen08,Taylor08,Maze08,Balasubramanian08,Gijs10b} and photonic devices \cite{Fu08,Babinec10,Hadden10,Toeno10,Siyushev10,Englund10}.
However, despite the rapid experimental progress the understanding of the optically-induced spin-dynamics of the NV centre is still incomplete. In particular, the spin-dependent intersystem crossing (ISC) rates as well as the number of singlet states involved in the optical cycle are still debated. These parameters are responsible for optical spin initialization and readout, and are important for a correct estimation of photon emission rates. We extract these values by a series of room-temperature experiments, where we perform spin-resolved fluorescence lifetime measurements using picosecond optical excitation pulses. The lifetime of the singlet manifold is measured by analyzing the initial fluorescence rate for consecutive microsecond optical pulses with variable delay. The temperature dependence of this lifetime yields insight into the number of involved singlet states. 

\section{Experimental setting}
We investigate individual NV centres contained in a high-temperature high-pressure (HTHP) grown type IIa diamond sample from Element Six ($\left\langle 111 \right\rangle$ oriented). The sample is studied in a scanning confocal microscope setup operated at $T=10 \ldots 300$\,K. Spin control is achieved via microwave (MW) fields applied to an Au-waveguide that is lithographically defined on the diamond surface \cite{Fuchs09,Gijs}. For optical excitation we use a continuous-wave (CW) laser at $\lambda = 532$\,nm, equipped with an acousto-optic modulator (AOM) with 20\,ns rise-time, as well as a frequency-doubled diode laser at $\lambda = 532$\,nm with a pulse length of 62\,ps (max. pulse energy: 25\,nJ) and variable repetition rate. For photon detection we use an avalanche photo diode in the single-photon counting regime with a timing jitter of 450\,ps. Time-resolved data are acquired using a time-correlated single photon counting module with a jitter of 10\,ps, using a bin size of 512\,ps. An arbitrary waveform generator (channel-to-channel jitter $<100$\,ps) is used as timing source of the experiment. 

\section{Model}
\begin{figure}
  \centering
  \includegraphics{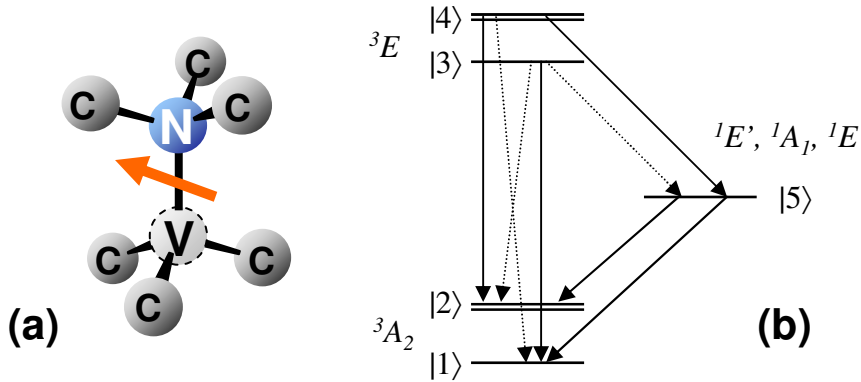}
  \caption{\label{fig:fig1} {\bf (a)} Lattice structure of the NV centre: a substitutional nitrogen atom {\sffamily N} next to a vacancy {\sffamily V} in the diamond lattice {\sffamily C}. {\bf (b)} Level structure of the NV centre: we consider spin-conserving ($k_{31}$, $k_{42}$) and spin-flip ($k_{32}$, $k_{41}$) transitions between triplets (states with spin projection $m_s = \pm 1$ are merged into states $\left|2\right\rangle$, $\left|4\right\rangle$). Spin-dependent ISC rates connect triplets to the singlet states (summarized as $\left|5\right\rangle$, and described by rates $k_{35}, k_{45}, k_{51}, k_{52}$).}
\end{figure}

The photo-dynamics of the NV centre [figure~\ref{fig:fig1}(a)] are determined by six electrons, which in the ground state form a triplet occupying an orbital of $^3A_2$ symmetry. The centre can be excited via a dipole-allowed transition to a $^3E$ triplet state. This level also has a spin-dependent probability to undergo ISC to a series of singlet states \cite{Manson06a}. We use a five-level model to describe the spin dynamics of the NV centre [figure~\ref{fig:fig1}(b)]. Spin-spin interaction splits the $^3A_2$ ground state by $D_{GS}=2.87$\,GHz into a state with spin projection $m_s=0$ ($\left|1\right\rangle$) and a doublet with $m_s=\pm1$ (summarized in $\left|2\right\rangle$). Correspondingly, the excited $^3E$ state is labelled $\left|3\right\rangle$ ($m_s=0$, associated with a lifetime $T_{1,3}$) and $\left|4\right\rangle$ ($m_s=\pm1$, lifetime $T_{1,4}$), and split by $D_{ES}=1.43$\,GHz \cite{Fuchs08,Neumann}. 
Two singlet states with a splitting of $\Delta E = 1.189$\,eV have been identified experimentally \cite{Manson08,Acosta}, but recent theoretical studies \cite{Gali10} and also data obtained in this work suggest the presence of a third singlet state between $^3A_2$ and $^3E$. For the analysis of spin dynamics, however, we summarize the singlet states in $\left|5\right\rangle$, and the corresponding lifetimes are summed and denoted as $T_{1,5}$. Rates from state $\left|m\right\rangle$ to state $\left|n\right\rangle$ are denoted by $k_{mn}$, and we only consider relaxation rates indicated in figure~\ref{fig:fig1}(b). Population in state $\left|n\right\rangle$ is denoted by $P_n$, and the spin polarization in ground and excited state are denoted by $P_{GS}=P_1/(P_1+P_2)$ and $P_{ES}=P_3/(P_3+P_4)$.

\section{Spin-dependent lifetime}
Pulsed optical excitation and time-resolved detection of fluorescence provides a simple and direct way to determine the lifetime of the excited state in an optical transition. If the excitation pulse is short compared to the lifetime $T_1$, the detected time-resolved fluorescence (averaged over many excitation cycles) decays exponentially $I \propto \exp(-t/T_1)$. However, if the system under consideration is excited into a mixture of $n$ excited states with different lifetimes, the detected fluorescence decays according to a multi-exponential function $ I \propto \sum_n a_n \exp(-t/T_{1,n})$. 

This situation is present in the case of NV centres in diamond, where the excited state is composed of a spin triplet. Population in this state decays radiatively to the triplet ground state with identical oscillator strength for the different spin projections \cite{Batalov}. Because of spin-dependent ISC \cite{Manson06a}, the effective lifetime of the excited state is significantly different for states $\left|3\right\rangle$ and $\left|4\right\rangle$ .
In literature \cite{Neumann, Batalov, Collins83} the NV centre's lifetime generally is obtained by fitting the time-resolved fluorescence to a single-exponential decay, leading to values of $T_{1,3}\approx 12 - 13$\,ns and $T_{1,4}\approx 8$\,ns in bulk diamond. Since optically induced spin polarization in NV centres is limited \cite{Manson06b, Fuchs10}, in fact we expect such a lifetime measurement to yield a bi-exponential decay curve with time constants set by the sums of rates out of states $\left|3\right\rangle$ and $\left|4\right\rangle$ [$T_{1,3}=1/(k_{31}+k_{32}+k_{35})$ and $T_{1,4}=1/(k_{41}+k_{42}+k_{45})$], and an amplitude ratio set by the initial spin polarization. Such a bi-exponential decay has been observed in \cite{Fuchs10}, where a polarization of $P_{ES}=0.84\pm0.08$\,\% has been obtained. These data are based on fluorescence lifetime measurements with MW spin manipulation in the excited state, where the duration of the MW pulse was neglected, and the pulse was assumed to be perfect. 

\begin{figure}
  \centering
  \includegraphics[width=15cm]{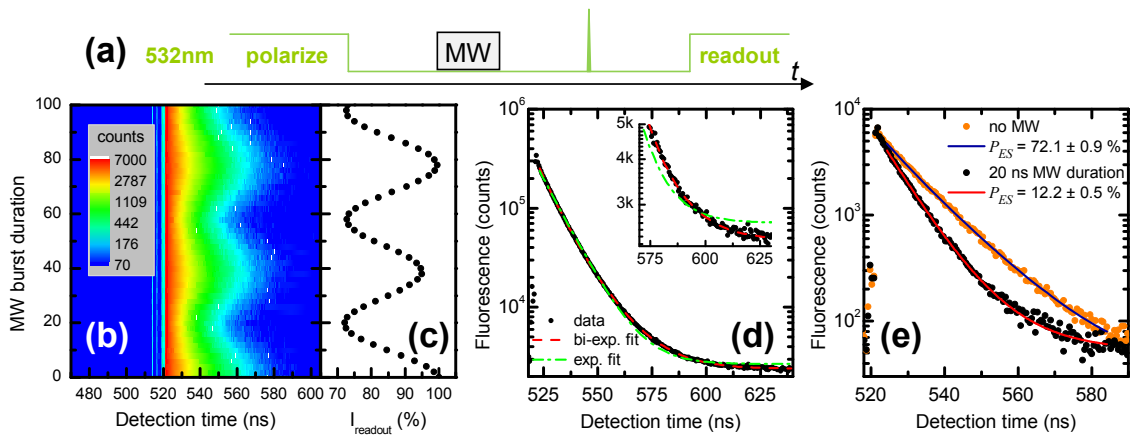}
  \caption{\label{fig:fig2} {\bf (a)} We first polarize the spin by applying a 1.3\,$\mu$s laser pulse at $\lambda=532$\,nm. After 800\,ns we turn on a MW field at 2.87\,GHz for a variable duration. 200\,ns after begin of the MW we excite the NV center by a 62\,ps laser pulse at $\lambda=532$\,nm and measure the time-resolved emission. The first 300\,ns of the subsequent polarization pulse is used for spin readout. The experiment was performed at $T=300$\,K. {\bf (b)} Fluorescence decay as function of MW burst duration (fluorescence counts are encoded in a logarithmic colour scale). The decay time oscillates with the $m_s = 0$ amplitude, as confirmed by {\bf (c)} conventional spin readout. {\bf (d)} Fluorescence decay curve, integrated over all applied MW burst durations. A fit using a bi-exponential function, yields the lifetime of states $\left|3\right\rangle$ and $\left|4\right\rangle$. A single-exponential function (shown for comparison) cannot accurately fit the experimental data. {\bf (e)} Degree of spin polarization $P_{ES} = P_{3}/(P_{3}+P_{4})$ with no MW applied and after a 20\,ns MW pulse, obtained from the relative amplitudes of a bi-exponential fit.}
\end{figure}

Here we present a simple and reliable way to obtain the spin-dependent lifetimes, without any assumptions on the MW pulse and spin polarization. For that purpose we drive Rabi oscillations in a conventional fashion, i.e. we apply a 1.3\,$\mu$\,s long off-resonant laser pulse to polarize the electron spin, followed by a MW pulse of variable duration, resonant with the zero-field splitting of $D = 2.87$\,GHz. After the MW, we apply a ps laser pulse. The fluorescence as function of MW pulse duration is shown in figure~\ref{fig:fig2}(b). For comparison we also plot the result of a conventional spin readout [figure~\ref{fig:fig2}(c)], i.e. the intensity of the first 300\,ns of the subsequent polarization laser pulse \cite{Jelezko04}. The data in figure~\ref{fig:fig2}(b) clearly reveal the oscillations in decay time which are in phase with oscillations of the electron spin. To accurately fit these data to a bi-exponential decay we sum over all MW pulse durations, which gives similar contribution from each spin state. From the fit we obtain the two time constants $T_{1,3} = 13.7 \pm 0.1$\,ns and $T_{1,4} = 7.3 \pm 0.1$\,ns [figure~\ref{fig:fig2}(d)]. By using these values as constants for a fit to the individual decay curves, we can determine the relative contributions of states $\left|3\right\rangle$ and $\left|4\right\rangle$ to the minima and maxima of the Rabi oscillation data from the relative amplitudes of the two exponentials. For this particular centre, we find $P_{ES,max}=72.1\pm0.9\%$ and $P_{ES,min}=12.2\pm0.5\%$ [figure~\ref{fig:fig2}(e)]. The value of $P_{GS,max}$ may be larger due to a nonßperfect spin conservation in optical excitation (see section \ref{sec:sec_pol_prob}).

We note that for a spin $S=1$ system with degenerate levels $m_s=\pm1$ (i.e. for NV centres in absence of magnetic field), the effect of resonant MW driving starting from a pure $\left|m_s=0\right\rangle$ state is to cause coherent oscillations between $\left|m_s=0\right\rangle$ and the symmetric superposition $\left|m_s=+1\right\rangle + \left|m_s=-1\right\rangle$. In a more realistic scenario we need to consider an only partially polarized state, where the density matrix subspace spanned by $\left|m_s=+1\right\rangle$ and $\left|m_s=-1\right\rangle$ has equal population of the symmetric and antisymmetric superposition state. Only the symmetric state will be transferred back to $\left|m_s=0\right\rangle$, so the effect of a half-oscillation is to transfer the full $m_s=0$ population into $m_s=\pm1$, while only half of the incoherent population in $m_s=-1$ and $m_s=+1$ is transferred back into $m_s=0$. As a consequence, when driving Rabi oscillations, the maximum population in $m_s=0$ will only reach half the value of the maximum population in $m_s=\pm1$ [figure~\ref{fig:fig2}(e)].

In summary, the presented method allows for accurate determination of the lifetime of the pure states $\left|3\right\rangle$ and $\left|4\right\rangle$, without the need of assumptions on the quality of spin manipulation. The knowledge of these lifetimes can then be used to quantify the spin polarization.

\section{Temperature dependence of singlet decay}
An important parameter for the spin dynamics of the NV centre under optical excitation is the time population spends in the singlet manifold before it decays back to the triplet ground state. This time scale is long compared to the excited state lifetime, and since decay into the singlet states (ISC) is favoured for $m_s=\pm1$, this leads to reduced fluorescence for $m_s=\pm1$, a fact which is routinely used for non-resonant spin readout \cite{Jelezko04}. 

\begin{figure}
  \centering
  \includegraphics{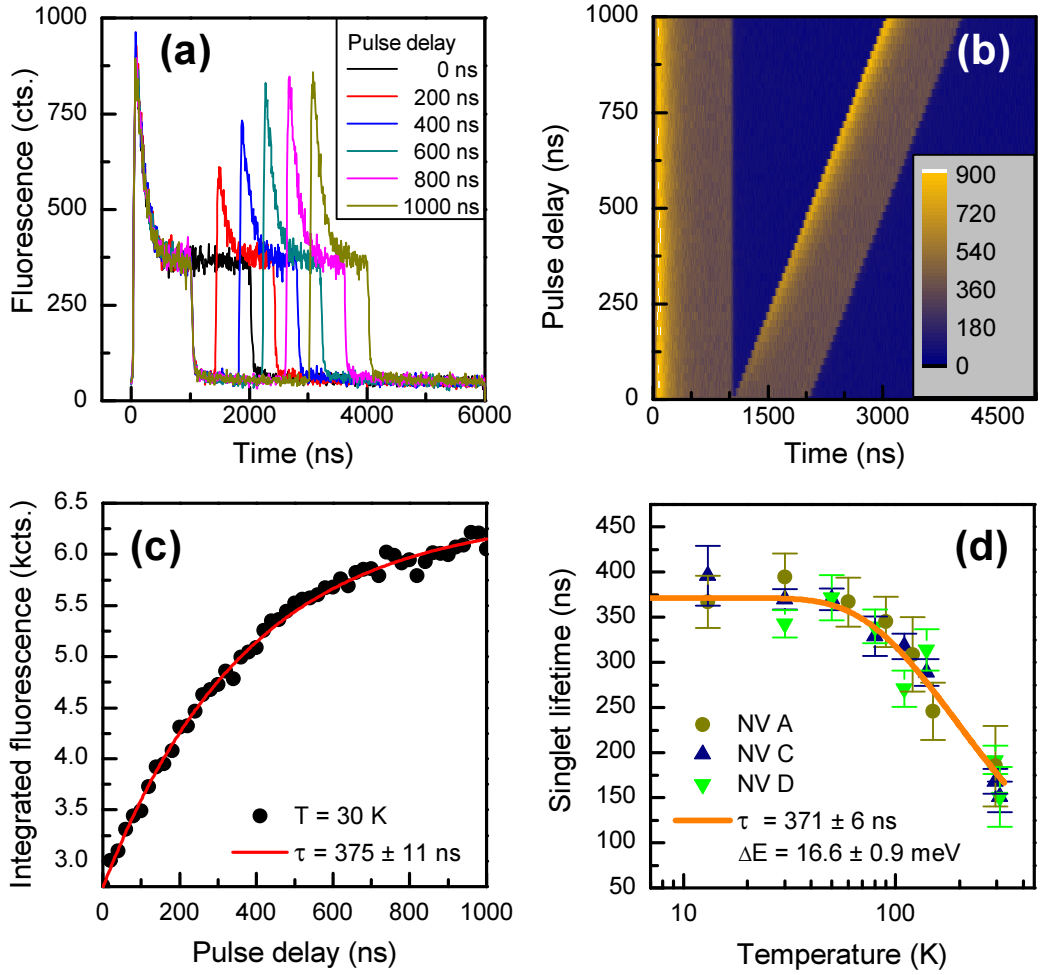}
  \caption{\label{fig:fig3} Decay of population from singlet states leads to recovery of fluorescence: {\bf(a)} individual traces and {\bf(b)} full data set of NV fluorescence with two excitation pulses and variable delay. {\bf(c)} Fluorescence counts integrated over first 30\,ns of second pulse as function of inter-pulse delay. The exponential increase is caused by decay out of singlet states. {\bf(d)} Temperature dependence of singlet decay rate. The fit assumes a phonon-assisted decay process.}
\end{figure}

The temperature dependence of this decay rate yields information about the energy splitting involved in this relaxation process, and adds further evidence for the number of singlet states contributing to the optical cycle of NV centres. Recently an infrared (IR) emission channel was observed, and attributed to a dipole-allowed transition between two singlet levels with an energy splitting of $\Delta E = 1.189$\,eV  \cite{Manson08, Acosta}. In the same publications it was shown that the IR emission follows the same time dependence as the visible transition, implying a short lifetime of the upper singlet state. Therefore the previously observed long-lived singlet state was attributed to the ground state of this IR transition. 
The temperature dependence of this singlet state lifetime has been reported for an ensemble of NV centres based on IR absorption measurements \cite{Acosta}. Here, we present data obtained on single NV centres. Since the oscillator strength of the IR transition is very weak, we use an indirect way to obtain this timescale, following the method used in \cite{Manson06a}: We first apply a green pulse (1\,$\mu$s) to reach a steady state population in the singlet states. This population manifests itself in reduced steady-state fluorescence with respect to the beginning of the pulse. After a variable delay we apply a second 1\,$\mu$s pulse [figure~\ref{fig:fig3}(a,b)]. During the delay between the pulses, population stored in the singlet states decays back to the triplet ground state. The fluorescence at the beginning of the second pulse is proportional to the population in the triplet ground state and thus we can attribute its dependence on inter-pulse delay to the population decay out of the singlet state. In figure~\ref{fig:fig3}(c) we integrate the photons emitted during the first 30\,ns of the second pulse for each inter-pulse delay and fit these data to an exponential function. Figure~\ref{fig:fig3}(d) summarizes the timescales we obtained in this way, for temperatures ranging from $T=13$\,K to $T=300$\,K for three different NV centres. 

We model the lifetime $\tau$ of the singlet state as a combination of temperature-independent spontaneous decay rate $\tau^{-1}_{0}$ and a rate accounting for stimulated emission of phonons of energy $\Delta E$ with an occupation given by Bose-Einstein statistics: $\tau = \tau_0 [1-\exp(-\Delta E/k_BT)]$. The fit yields a spontaneous emission lifetime of $\tau = 371 \pm 6$\,ns and a phonon energy of $\Delta E = 16.6 \pm 0.9 $\,meV, in reasonable agreement with ensemble data obtained in \cite{Acosta}. 

This result suggests that, apart from the 1.189\,eV splitting, a third level $\Delta$E below the IR transition's ground state is involved in the optical cycle of the NV centre. We can exclude that this third level is the triplet ground state, just 16.6\,meV below the lowest singlet, as this would imply phonon-assisted spin relaxation on a sub-microsecond time scale at room temperature. This scenario clearly contradicts experimentally observed spin lifetimes on a millisecond timescale \cite{Neumann08,Gijs}. Consequently the third level is likely to be another singlet state. The presence of three singlet states in between the triplet $^3A_2$ and $^3E$ states was recently predicted by an ab-initio calculation of the excited states in the NV centre \cite{Gali10}, however, there a larger energy splitting between the lowest singlet states was obtained. 

\section{\label{sec:sec_pol_prob}Polarization probability}
Polarizing the electron spin by off-resonant optical excitation is a key technique for room-temperature spin manipulation of NV centres. Although this effect already has been identified to be caused by a spin-dependent ISC rate \cite{Manson06a, Nizovtsev03, Harrison04, Nizovtsev05}, little is known about the relative contributions of spin-flip transitions between triplet states ($k_{32}$, $k_{41}$), and ISC rates ($k_{35}$, $k_{45}, k_{51}$, $k_{52}$). We address this question by determining the polarization change due to a single excitation cycle. 

\begin{figure}
  \centering
  \includegraphics{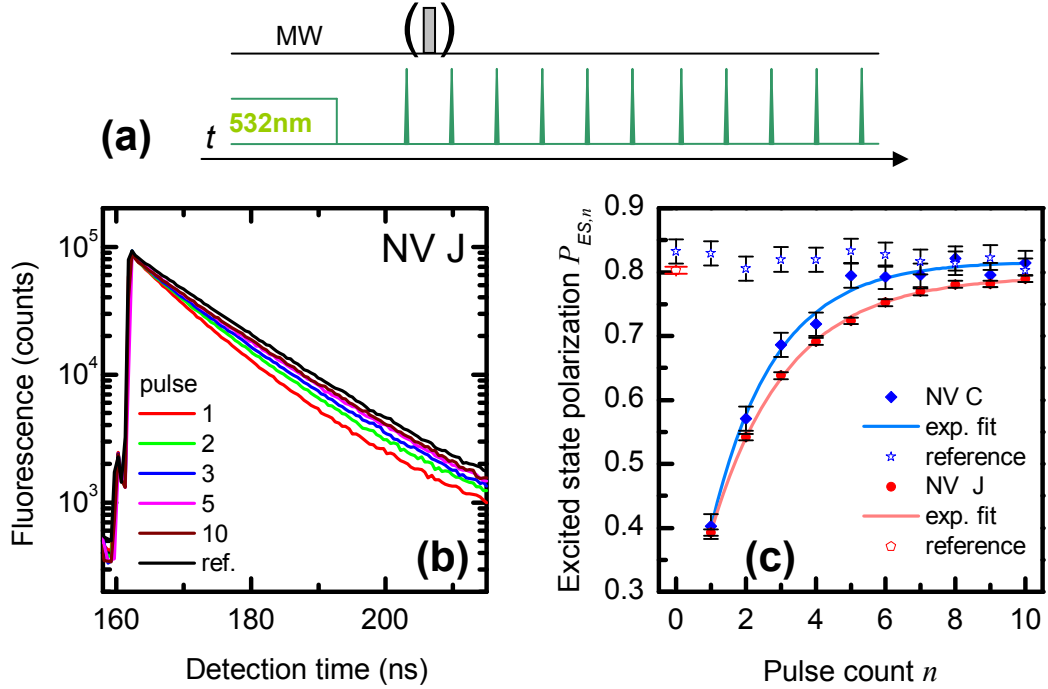}
  \caption{\label{fig:fig4} {\bf (a)} We first polarize the spin by means of a 2\,$\mu$s laser pulse at $\lambda=532$\,nm. After a delay of 1\,$\mu$s 
  we apply a sequence of 11 pulses of 62\,ps duration at $\lambda=532$\,nm. For NV J, we apply a MW pulse 1\,$\mu$s after the first ps-pulse to invert the spin state (for NV C we alternately run sequences with and without MW pulse). For each ps-pulse we measure the time-resolved emission. The experiment was performed at $T=300$\,K. {\bf(b)} Fluorescence decay for consecutive ps-excitation pulses. The decay follows a bi-exponential function. A reference ps-pulse after a green 2 $\mu s$ spin polarization pulse yields the initial optically induced spin polarization. A subsequent MW pulse transfers polarization to $\left|2\right\rangle$. {\bf(c)} Change in polarization between consecutive pulses yields spin-flip probabilities $p_{12}(\left|1\right\rangle \rightarrow \left|2\right\rangle)$ and $p_{21}(\left|2\right\rangle \rightarrow \left|1\right\rangle)$ per optical excitation cycle.}
\end{figure}

For that purpose we first initialize the NV spin by a 2\,$\mu$s polarization pulse. After a waiting time of 1\,$\mu$s we excite the NV centre by a reference ps-pulse and measure the spin polarization by analyzing the relative contributions of the amplitudes in a bi-exponential fit to the fluorescence decay curve, as outlined in the previous section. This polarization corresponds to the steady-state value after CW excitation. To determine the change in polarization per excitation cycle we now apply a MW pulse to transfer the $m_s=0$ population into the $m_s=\pm1$ states and then drive individual excitation cycles by applying 10 consecutive ps-pulses separated by 2\,$\mu$s. For each excitation cycle, we again determine the spin polarization [figure~\ref{fig:fig4}(c)]. All these bi-exponential fits use the same two time constants, obtained from a fit to the sum of all decay curves. From a power dependence measurement of the NV fluorescence rate we determine an excitation probablity $\alpha=0.95\pm0.05$.

The effect of a single excitation cycle on the spin polarization can be described by two counter-acting probabilities: a spin-flip from $\left|1\right\rangle$ to $\left|2\right\rangle$ ($p_{12}$) and the opposite process ($p_{21}$). Here, we consider only optically induced effects, i.e. the time between excitation pulses $\Delta_t$ is assumed to be much shorter than the spin-lattice relaxation time (this assumption is substantiated by the constant polarization $P_{ES}$ for the 10 consecutive reference pulses in case of NV C [figure~\ref{fig:fig4}(c)]). For state $\left|m\right\rangle$, population just before pulse $n$ is denoted by $P_{m,n}$, and just after the excitation pulse by $P'_{m,n}$. The steady state value $P_{m,n=\infty}$ is abbreviated by $P_m$. The pulse separation $\Delta_t$ is much larger than the singlet decay time, such that $P_{1,n}+P_{2,n}=1$ and therefore $P_{GS,n} = P_{1,n}$. 

Experimentally, we obtain spin polarization in the excited state. This differs from the ground state polarization due to a fraction $\epsilon = k_{23} / k_{13} = k_{14}/k_{24}$ of spin non-conserving transitions. The populations for the $(n+1)^{th}$ excitation pulse are then given by:

\begin{eqnarray}
	P_{1,n+1} &= \alpha p_{21}P_{2,n} + (1 - \alpha p_{12})P_{1,n} \\
	P_{2,n+1} &= \alpha p_{12}P_{1,n} + (1 - \alpha p_{21})P_{2,n} \\
	P'_{3,n+1} &= \alpha \left(\frac{\epsilon}{1+\epsilon}P_{2,n+1}+\frac{1}{1+\epsilon}P_{1,n+1}\right) \\
	P'_{4,n+1} &= \alpha \left(\frac{\epsilon}{1+\epsilon}P_{1,n+1}+\frac{1}{1+\epsilon}P_{2,n+1}\right).
\end{eqnarray}

The asymptotic value of the polarization is $P_{GS}=p_{21}/(p_{12}+p_{21})$, and the excited state $P_{ES,n}=P'_{3,n}/(P'_{3,n}+P'_{4,n})=\left[P_{1,n}\left(1-\epsilon\right)+\epsilon\right]/\left(1+\epsilon\right) \approx P_{1,n}$ closely follows the ground state spin polarization for small $\epsilon$. From \cite{Manson06a} we can obtain an upper bound of $\epsilon\leq0.04$. The change in polarization per pulse [figure~\ref{fig:fig4}(c)] can be fitted by an exponential function $P_{ES}(n)=P_{ES}+a\exp(-n/c)$. From the steady-state polarization $P_{ES}$ and the polarization rate $c$ we can extract the spin-flip probabilities $p_{21}$ and $p_{12}$ per optical cycle. These probabilities depend only weakly on $\epsilon$. Results are summarized in table \ref{tab:tab1}:

\begin{table}[h]
\caption{\label{tab:tab1} Summary of parameters ($T=300$\,K), taking $\alpha=0.95\pm0.05$ and $\epsilon = 0.01\pm0.01$. $p_{12}$ and $p_{21}$ are the total spin-flip probabilities per optical cycle. $p_{35}$, $p_{45}$, $p_{51}$ and $p_{52}$ are the spin-dependent ISC probabilities for population in state $\left|3\right\rangle$, $\left|4\right\rangle$ and $\left|5\right\rangle$.}

\begin{indented}
\item[]\begin{tabular}{l||r|r}
  & NV J & NV C \\
\br
$p_{12}$ & 0.078 $\pm$ 0.002 & 0.079 $\pm$ 0.004 \\
$p_{21}$ & 0.315 $\pm$ 0.011 & 0.372 $\pm$ 0.017 \\
\mr
$T_{1,3} (ns)$ & 13.26 $\pm$ 0.03 & 13.1 $\pm$ 0.1\\
$T_{1,4} (ns)$ &  6.89 $\pm$ 0.06 &  7.0 $\pm$ 0.2\\

$p_{35}$ & 0.14 $\pm$ 0.02 & 0.17 $\pm$ 0.03 \\
$p_{45}$ & 0.55 $\pm$ 0.01 & 0.56 $\pm$ 0.02 \\
$p_{51}/p_{52}$ & 1.15 $\pm$ 0.05 & 1.6 $\pm$ 0.4 

\end{tabular}

\end{indented}
\end{table}

\begin{figure}
  \centering
  \includegraphics{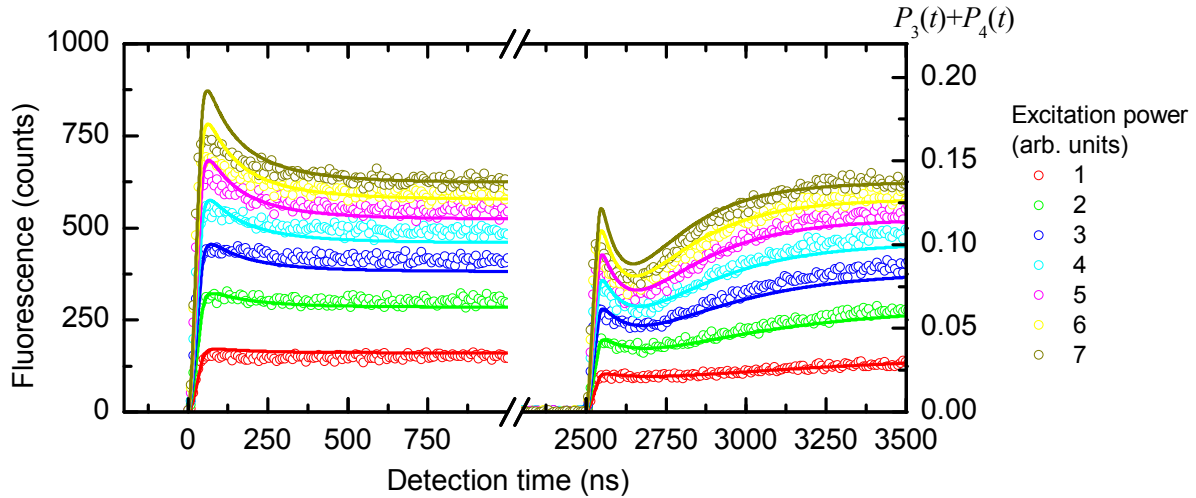}
  \caption{\label{fig:fig5} 
  Time-resolved excitation-power dependence of NV fluorescence excited by off-resonant 1\,$\mu$s pulses (experimental data and model). The first pulse displays the fluorescence of NV centres that are optically polarized into $m_s=0$. The second pulse is obtained after applying a MW $\pi$-pulse, representing the fluorescence of a NV centre polarized in $m_s=\pm1$. The model shows the population in states $\left|3\right\rangle$ and $\left|4\right\rangle$. The calculations are based on parameters from NV J, assuming $\epsilon=0.01$ and $\alpha=1$ (no fit is applied). Excitation rates are taken as integer multiples of 4 MHz.}
\end{figure}

If we assume only rates indicated in figure~\ref{fig:fig1}(b), use experimentally obtained lifetimes $T_{1,3}$, $T_{1,4}$ and take $\epsilon=0.01\pm0.01$ and $\alpha=0.95\pm0.05$, we can obtain the spin-dependent ISC probabilities of $p_{45}=k_{45}/(k_{41}+k_{42}+k_{45})$ and $p_{35}=k_{35}/(k_{31}+k_{32}+k_{35})$, and for the reverse process $p_{51}/p_{52}=k_{51}/k_{52}$, as summarized in table \ref{tab:tab1}. The low values of $p_{51}/p_{52}$ are in contrast to the current picture of the singlet relaxation process, where $p_{52} = 0$ was used \cite{Manson06a}. This gives another hint that an additional $^1E$ state needs to be considered for the relaxation process. The temperature dependence of $p_{51}/p_{52}$ could give further insight into this topic. 

We note that this set of parameters can be used to model the spin- and power dependence of time-resolved NV centre emission using $\mu$s excitation pulses at $\lambda=532$\,nm within the framework of this five-level model, with reasonable agreement to experimental data [figure~\ref{fig:fig5}].

\section{Summary}
We have experimentally determined the spin-dependent lifetime of the NV centre's excited state, whose difference is dominated by a spin-dependent ISC rate. Knowledge of these lifetimes allows us to determine the degree of spin polarization. In a second experiment, we identified the total lifetime of the singlet states, and, by analyzing its temperature dependence, the energy splitting of the long lived singlet transition. The measured energy of $\approx16$\,meV indicates that at least three singlet states are involved in the optical cycle of the NV centre. Finally, we determined the spin-dependent ISC probabilities by analyzing the change of spin polarization induced by a single excitation cycle, without making assumptions on number and nature of the singlet states. The obtained rates are consistent with spin-dependent NV fluorescence dynamics based on a five-level model. 

\ack
We would like to thank V. V. Dobrovitski and L. Childress for helpful discussions and Daniel Twitchen and Paul Balog of Element Six for provision of the diamond sample. This work is supported by the Dutch Organization for Fundamental Research on Matter (FOM), the Netherlands Organization for Scientific Research (NWO) and the EU SOLID programme. L. R. acknowledges support of the European Community under a Marie-Curie IEF fellowship.

\end{document}